\pdfoutput=1
\documentclass[english]{article}
\usepackage{geometry}
\usepackage{babel}
\usepackage{xcolor}
\usepackage{graphicx}
\usepackage{amsmath,amsfonts}
\usepackage{xcolor}
\usepackage{graphicx,indentfirst,subfigure,epsfig}
 \usepackage{varioref}%  smart page, figure, table, and equation referencing
 \usepackage{wrapfig}%   wrap figures/tables in text (i.e., Di Vinci style)
 \usepackage{subfigure}% subcaptions for subfigures
 \usepackage{subfigmat}% matrices of similar subfigures, aka small mulitples
 \usepackage{amsthm}
\usepackage{bm}

  \def\clap#1{\hbox to 0pt{\hss#1\hss}}

\providecommand{\mat}[1]{\bm{#1}}%
\renewcommand{\vec}[1]{\mathbf{#1}}

\providecommand{\mA}{\ensuremath{\mat{A}}}
\providecommand{\mB}{\ensuremath{\mat{B}}}

\providecommand{\mI}{\ensuremath{\mat{I}}}

\providecommand{\mP}{\ensuremath{\mat{P}}}
\providecommand{\mQ}{\ensuremath{\mat{Q}}}
\providecommand{\mR}{\ensuremath{\mat{R}}}

\providecommand{\mU}{\ensuremath{\mat{U}}}
\providecommand{\mV}{\ensuremath{\mat{V}}}

\providecommand{\mX}{\ensuremath{\mat{X}}}

\providecommand{\va}{\ensuremath{\vec{a}}}
\providecommand{\vb}{\ensuremath{\vec{b}}}

\providecommand{\vr}{\ensuremath{\vec{r}}}

\providecommand{\vv}{\ensuremath{\vec{v}}}

\providecommand{\vx}{\ensuremath{\vec{x}}}
\providecommand{\vy}{\ensuremath{\vec{y}}}

\usepackage{hyperref}
\hypersetup{colorlinks, citecolor=red, linkcolor=blue, urlcolor=red, breaklinks=true}
\usepackage{scrextend}
\usepackage[T1]{fontenc}
\usepackage[utf8]{inputenc}
\usepackage{tgtermes}
\usepackage{nomencl}
\usepackage{algorithm}
\usepackage{algorithmic}
\usepackage{booktabs}
\usepackage{longtable}
\usepackage{lineno}
\makenomenclature

\begin{document}
\large
\title{\textbf{Spatial Flow-Field Approximation Using \\Few Thermodynamic Measurements \\ Part I: Formulation and Area Averaging}}
\author{Pranay Seshadri\footnote{Address all correspondence to \texttt{ps583@cam.ac.uk}}$\;^{\; \star \dagger}$, Duncan Simpson$^{\ddagger}$, George Thorne$^{\ddagger}$, \\Andrew Duncan$^{\smallint \dagger}$,  Geoffrey Parks$^{\star}$\\ \vspace{0.3 cm} \\ $^{\star}$Department of Engineering, University of Cambridge, Cambridge, U. K., \\ $^{\dagger}$Data-Centric Engineering, The Alan Turing Institute, London, U. K., \\ $^{\ddagger}$Civil Aerospace, Rolls-Royce plc., Derby, U. K., \\$^{\smallint}$Department of Mathematics, Imperial College, London, U. K.}
\date{}
\maketitle{}
%%\linenumbers
%%%%%%%%%%%%%%%%%%%%%%%%%%%%%%%%%%%%%%%%%%%%%%%%%%%%%%%%%%%%%%%%%%%%%%
\begin{abstract}
Our investigation raises an important question that is of relevance to the wider turbomachinery community: how do we estimate the spatial average of a flow quantity given finite (and sparse) measurements? This paper seeks to advance efforts to answer this question rigorously. In this paper, we develop a regularized multivariate linear regression framework for studying engine temperature measurements. As part of this investigation, we study the temperature measurements obtained from the same axial plane across five different engines yielding a total of 82 data-sets. The five different engines have similar architectures and therefore similar temperature spatial harmonics are expected. Our problem is to estimate the spatial field in engine temperature given a few measurements obtained from thermocouples positioned on a set of rakes. Our motivation for doing so is to understand key engine temperature modes that cannot be captured in a rig or in computational simulations, as the cause of these modes may not be replicated in these simpler environments. To this end, we develop a multivariate linear least squares model with Tikhonov regularization to estimate the 2D temperature spatial field. Our model uses a Fourier expansion in the circumferential direction and a quadratic polynomial expansion in the radial direction. One important component of our modeling framework is the selection of model parameters, i.e.~the harmonics in the circumferential direction. A training-testing paradigm is proposed and applied to quantify the harmonics. 
\end{abstract}

\section{Introduction and motivation}
\label{sec:intro}
It is virtually impossible today for a computational simulation to replicate the unsteady, complex, and random nature of fluid moving through an engine at cruise; there are simply far too many unknowns. Blades in a single row are susceptible to manufacturing variations, the casing is never perfectly circular, operating conditions exhibit some variability, and state-of-the-art 3D unsteady solvers model the turbulence and do not resolve all the pertinent scales and their evolution. Our best window into this world is undoubtedly the data that arises from engine tests.

The challenge with engine tests is that they typically have low instrumentation coverage as there is limited space for instrumentation and capacity for acquiring signals from the various measurement devices. Furthermore, access to specific areas of interest in the gas path is extremely limited. In a rig, by contrast, the temperatures, pressures and vibration levels will generally be reduced, though they do offer greater access for instrumentation. The measurement methods in a rig may also be different from those in an engine, i.e., certain probe designs that are not permissible in an engine may be used in a rig. That said, the rig is never fully representative of the engine, in particular, it does not fully capture the interactions between adjacent components in the engine. 

\begin{figure}
\centering
\includegraphics[width=0.8\textwidth]{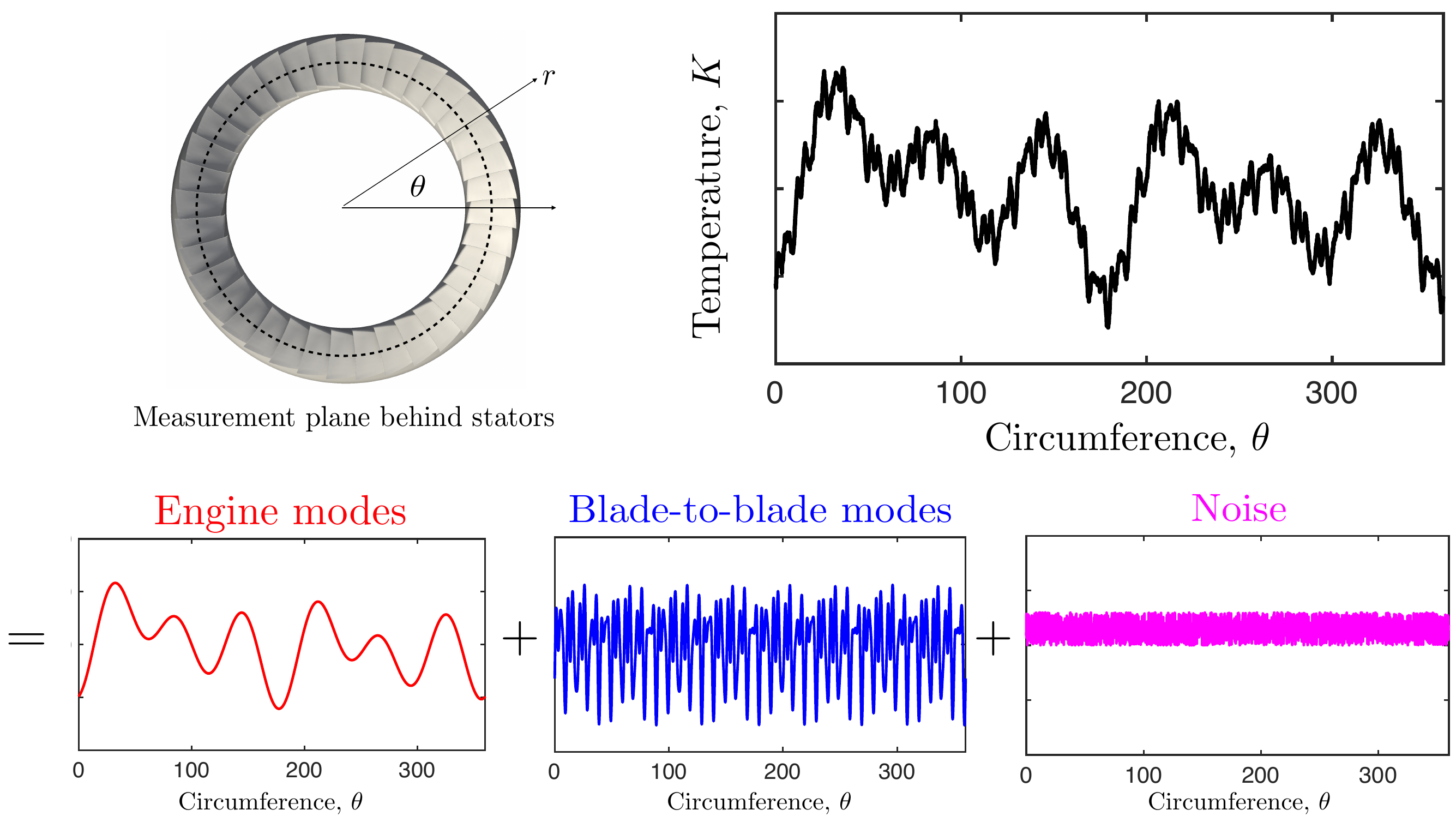}
\caption{Breakdown of engine temperature measurements.}
\label{fig:measurements}
\end{figure}

Our focus in this paper will be on on-ground engine temperature measurements, obtained from a series of rakes at a specific axial station. We wish to study these temperature values, as they cannot be reproduced in rigs or CFD. Consider the schematic shown in Figure~\ref{fig:measurements}. One can consider the spatial variation of engine temperature to be a superposition of engine modes (not visible in a rig), blade-to-blade modes (likely visible in a rig), and noise. The blade-to-blade modes originating from leakage flows, tip vortices, stator and rotor wakes and upstream potential fields lie at the higher end of the frequency spectrum (see page 9 of \cite{ernst2011analysis}) as these will be functions of the blade numbers. They can be experimentally determined using time resolved temperature and pressure traverse measurements. Annular asymmetries in these measurements are expected as a superposition of blade-to-blade modes with the engine modes will result in variations in viscous mixing effects, inviscid wake stretching and the transport of low momentum wake air across the passage \cite{sanders2002multi, mailach2008periodical}. Engine modes, by contrast, are captured by positioning rakes at the same pitch relative to the upstream stator vanes. Asymmetries in engine modes can be introduced by upstream bleed positions, upstream ducts, structural members and downstream potential fields. 

So what is typically done with engine measurements? The first notion is to arithmetically average them. In fact, historically, engines were fitted with equal area weighted probes (see \cite{stoll1979effect, francis1989measurement}) permitting relatively straightforward area average calculations. Over the years different averaging techniques have emerged (see Cumpsty and Horlock \cite{cumpsty2006averaging} for a thorough review) including the work-average method of Pianko \cite{pianko1983propulsion}, the momentum mixing method of Dzung \cite{dzung1971} and the mass average---all three requiring details on other primal flow quantities compared to the area average. Other notions include analyzing radial profiles and using them for engine prognostic and diagnostic efforts. 

But how precise is our averaging process given that we do not approximate the spatial flow field itself? Furthermore,  how do we account for the uncertainties introduced by both individual measurements and any spatial model? As engine measurements shape our understanding of the aerothermodynamics of the gas path, the answers to these questions are of great importance. Moreover, as these measurements (and subsequent inferences) make their way into preliminary design tools and overall uncertainty budgets, inaccurate estimates of both spatial approximations and their averages can have repercussions on new engine programs, measurement and testing practices, and even sub-system-level design.  

In the first part of this two-part paper (see part II in \cite{seshadri2019b}), we develop a framework for studying engine temperature measurements with a more principled, data-centric approach. Specifically, we propose a regression-based model for approximating the spatial flow-field at a given axial plane. To the best of the authors' knowledge, there remains a dearth of publicly available information on this subject; in section~\ref{sec:lit} we discuss prior work done in this area and highlight our contributions. In section~\ref{sec:model} we detail our regression-based model and deploy it on a range of engine data-sets. Finally, in section~\ref{sec:area}, we address the issue of area averaging. 

\mbox{}
 
\nomenclature[06]{$+a$}{Operator}
\nomenclature[03]{$2a$}{Number}
\nomenclature[05]{$:a$}{Punctuation Symbol}
\nomenclature[04]{$Aa$}{Upper Case Letter}
\nomenclature[01]{$aa$}{Lower Case Letter}
\nomenclature[02]{$\alpha$}{Greek Character}
 
\printnomenclature

\printnomenclature

\section{Data-centric spatial flow-field approximations}
\label{sec:lit}
In this section, we outline prior work undertaken from both the modeling and spatial averaging perspectives.

\subsection{Spatial modeling and area averaging}
As mentioned previously, there has been very little published on \emph{spatial approximation}, compared to  \emph{spatial averaging} (e.g.,  the aforementioned methods of Pianko and Dzung). One can broadly classify existing area averaging practices into three categories:
\begin{enumerate}
\item Numeric average: An ensemble average of all the measurements that does not factor into account radial and circumferential locations of the measurements (see page 28 in \cite{skiles1980turbine}). 
\item Area weighted: Each measurement is weighted by the sector area coverage associated with its probe (see page 8 in \cite{skiles1980turbine}). 
\item Fourier 1D: Temperature values along individual rakes are averaged radially (numeric average) and then a Fourier expansion is fitted circumferentially, typically using least squares (for the latter see page 9 of \cite{chilla2019reducing}). 
\end{enumerate}
The latter reference is noteworthy as the authors do make use of a least squares strategy to estimate the flow-field at a single radial height. 

That said, each of these averaging practices introduces uncertainties that are difficult to accurately quantify. We address the issue of uncertainties in spatial averages and the impact of imprecise measurements in our companion paper \cite{seshadri2019b}, and for the remainder of this paper focus solely on obtaining a spatial approximation of the flow-field, which we can then integrate analytically to obtain an area average.  

\subsection{Supervised machine learning}
There are numerous techniques that fall under the moniker of \emph{supervised learning}. They include linear regression, support vector machines, decision trees, linear discriminant analysis, principal component analysis and neural networks. Foundational texts that delve into techniques for supervised learning include Hastie, Tibshirani and Friedman \cite{friedman2009elements}, Rogers and Girolami \cite{rogers2016first} and more recently Goodfellow et al.~\cite{goodfellow2016deep}. In a nutshell, these methods are usually \emph{trained} on a set of data to establish values of the key model parameters. The model itself is typically associated with some \emph{basis}, e.g.~Fourier, polynomial, Gaussian kernels, etc. Establishing the values of these parameters involves optimizing an objective function, which can be defined either with respect to the $l_1$ or $l_2$ norm or some combination thereof. Once these model parameters have been established, the model is applied to a \emph{testing} data-set to certify its utility. 

The question we are faced with in this paper concerns how we obtain a 2D spatial representation using a few scattered measurements. To this end, conventional nearest neighbor or linear interpolation techniques fail to embed frequency information pertaining to key engine harmonics. Other techniques based on standard fast Fourier decompositions  demand far more rakes than can physically be placed in an engine---even for identifying a few key harmonics---owing to the sampling constraints imposed by the Nyquist-Shannon bound. Methods based on $l_1$ norm regression, e.g.~basis pursuit and least absolute shrinkage and selection operator (see Chapter 3 in \cite{friedman2009elements}), are contingent on sparsity in the model coefficients, a characteristic that is not necessarily guaranteed. Methods centered on $l_2$ norm regression are, in practice, less robust than $l_1$ approaches, but are useful in the absence of sparsity in the coefficients. Furthermore, they can be used even with a few measurements, a requirement in our work. 

\section{Data-driven spatial model}
\label{sec:model}
In this section we describe our approach to fit a parametric 2D model to the engine data. A key component of this modeling paradigm lies in our strategy for determining the harmonics.

\subsection{The data-sets}
First, we detail the data-sets that will be used throughout this work. We study the temperature measurements obtained from 5 different engines from the same engine family, at the same measurement plane, relative to its upstream and downstream components. The rakes in all the engines are axially placed downstream of a series of outlet guide vanes and are circumferentially positioned to avoid the wakes associated with these vanes. Details of the rake positions in the engines and the number of different extracts are provided in Table \ref{table:data-sets}. Each extract corresponds to a temperature measurement reading at a given engine power setting. The greater the power setting, the higher the mean temperature. Each measurement is obtained by sampling the raw thermocouple voltage at 192 kHz with a rolling average for the last 20 milliseconds. This signal is then filtered to remove noise components at 50 Hz and 400 Hz. Finally, the signal is averaged over a 30 second interval at a sampling rate of 33 Hz. Appropriate calibrations are then applied to convert this voltage into a temperature (in Kelvin) value. In this paper we will assume the uncertainties associated with this temporal averaging are negligible.
\begin{table}
\begin{center}
\caption{Description of the data-sets used.}
\begin{tabular}{|c|c|c|c|}
  \hline		
\textbf{Engine} &   \textbf{No. of rakes}  &   \textbf{Rake positions ($^\circ$) } &  \textbf{Extracts}  \\ \hline
 \hline
  A & 6 &  54.0, 90.0, 162.0, 234.0,  270.0, 342.0 & 13\\ \hline
  B & 6 & 54.0, 90.0, 162.0, 234.0,306.0, 342.0  & 20 \\ \hline
  C & 6 & 54.0, 90.0, 162.0, 234.0,  306.0, 342.0  & 26\\ \hline
  D & 6 & 54.0, 90.0, 162.0, 234.0, 306.0, 342.0 & 11\\ \hline
  E & 8 &  18.75,   60.625, 140.0, 179.58, 219.375,  258.75,  298.75, 340.0 & 12\\ 
 \hline  
\end{tabular}
\label{table:data-sets}
\end{center}
\end{table}

To permit rapid scripting of the measurements and subsequent analysis, the engine data was formatted into a series of XML documents and the Python \texttt{xml} library was used to extract the temperature data for subsequent modeling.  

\subsection{Multivariate regression model}
We now describe our modeling framework. We are given $N \times M$ temperature measurements, obtained from $N$ rakes and $M$ probes per rake at a given axial station in an engine. We define the measurement matrix $\mB \in \mathbb{R}^{N \times M}$
\begin{equation}
\mB=\left(\begin{array}{ccc}
\mid &  & \mid\\
\vb_{1} & \ldots & \vb_{M}\\
\mid &  & \mid
\end{array}\right)
\end{equation}
where $\vb_i \in \mathbb{R}^{N}$ for $i=1, \ldots, M$ is a vector of the temperature values for the $i^{\text{th}}$ rake. In this section, we will assume that there are no errors in $\mB$, and our aim here is to develop a model that best describes the spatial variation in temperature using the finite measurements in $\mB$. More specifically, we are trying to determine the whole-engine variation and not the blade-to-blade variation. Given the harmonic nature associated with this sector-to-sector variation, we will use a Fourier expansion in the circumferential direction. We define the Fourier matrix $\mA \in \mathbb{R}^{N \times \left( 2k+1 \right)}$ where $k$ represents the number of harmonics. Elements of $\mA$ are given by
\begin{equation}
a_{ij}=\begin{cases}
\begin{array}{cc}
1.0 & \text{if}\; \; j=1,\\
\sin\left(\omega_{\frac{j}{2}}\pi\theta_{i}/180^{\circ}\right) & \text{if} \; \; j>1\; \; \text{when} \;  j \; \text{is even},\\
\cos\left(\omega_{\frac{j-1}{2}}\pi\theta_{i}/180^{\circ}\right) & \text{if} \; \; j>1\; \; \text{when} \;  j \; \text{is odd},
\end{array}\end{cases}
\end{equation}
where $\boldsymbol{\theta} = \left(\theta_{1},\ldots,\theta_{N}\right)$ is the circumferential location of the $N$ rakes in degrees. The number of harmonics $k$ in $\boldsymbol{\omega}=\left(\omega_{1},\ldots,\omega_{k}\right)$ used in the spatial approximation is chosen such that $N > \left( 2k+1 \right)$. 

Our objective is to solve the \emph{multivariate regression} problem
\begin{align}
\begin{split}
\hat{\mX} &=\underset{\mX}{\text{argmin}}\left\Vert \mA\mX-\mB\right\Vert _{2}^{2} \\
&=\left( \mA^{T} \mA\right)^{-1} \mA^{T} \mB
\label{equ_lsqt}
\end{split}
\end{align}
where the columns of the unknown matrix $\mX \in \mathbb{R}^{ \left( 2k+1 \right) \times M }$ represent the Fourier coefficients at each of the radial positions associated with the $M$ probes. Solutions to \eqref{equ_lsqt} can be readily found provided that $\mA^{T} \mA$ is not singular. We remark here that \eqref{equ_lsqt} does not distinguish between measurements obtained at mid-span and those towards the hub or casing. One can, however, premultiply both $\mA$ and $\mB$ by a weight matrix that allocates greater preference to obtaining a model that is accurate at mid-span than at other spanwise locations. 

Consider a parameterized analogue of $\mA$, denoted by $\va \left(\theta \right) \in \mathbb{R}^{\left( 2k+1 \right)}$, where
\begin{align}
\va^{T} \left(\theta \right)  & = \left( a_1 \left( \theta \right) , \ldots, a_{2k + 1 }\left( \theta \right) \right) \;\; \; \text{with} \; \; \; \\
a_{j}\left( \theta \right) &= \begin{cases}
\begin{array}{cc}
1.0 & \text{if}\; \; j=1,\\
\sin\left(\omega_{\frac{j}{2}}\pi\theta/180^{\circ}\right) & \text{if} \; \; j>1\; \; \text{when} \;  j \; \text{is even},\\
\cos\left(\omega_{\frac{j-1}{2}}\pi\theta/180^{\circ}\right) & \text{if} \; \; j>1\; \; \text{when} \;  j \; \text{is odd},
\end{array}\end{cases}
\end{align}
It is clear that the product $\va^{T} \left(\theta \right) \mX$ corresponds to the temperature values at a specific value of $\theta$ across all $M$ radial locations. Now, to obtain a complete spatial representation of temperature, we need to radially interpolate these values using a polynomial with an appropriately selected degree. Once again, we resort to solving this as a least squares problem. We define the Vandermonde matrix $\mV \in \mathbb{R}^{M \times p}$ where $\left(p - 1 \right)$ is the highest degree of the polynomial. Columns of $\mV$ have the form
\begin{equation}
\mV=\left(\begin{array}{ccccc}
\mid & \mid & \mid &  & \mid\\
\boldsymbol{1} & \vr & \vr^{2} & \ldots & \vr^{p-1}\\
\mid & \mid & \mid &  & \mid
\end{array}\right)
\end{equation}
where $\vr= \left( r_1, \ldots, r_M \right)^{T} \in \mathbb{R}^{M}$ stores the radial locations of the $M$ taps. Then, we solve for the coefficients $\vy$ associated with the least squares problem
\begin{equation}
\hat{\vy}\left( \theta\right) =\underset{\vy}{\text{argmin}}\left\Vert \mV\vy-  \left(   \va^{T}\left(\theta\right)  \mX  \right)^{T} \right\Vert _{2}^{2} = \left(\mV^{T} \mV \right)^{-1} \mV^{T} \mX^{T} \va\left( \theta \right) .
\label{equ_lsqt_2}
\end{equation}
Let us now define a parameterized analogue of $\mV$, denoted by $\vv\left( r \right) \in \mathbb{R}^{p}$, where 
\begin{equation}
\vv^{T}\left(r\right)=\left(\begin{array}{ccccc}
1 & r & r^{2} & \ldots & r^{p-1}\end{array}\right).
\end{equation}
Thus the product $\vv^T \left(r\right) \vy \left( \theta\right)$ gives us the temperature approximation given the tuple $\left( r, \theta \right)$. For completeness we give the full expression
\begin{align}
\begin{split}
T\left(r,\theta\right) &=\vv^{T}\left(r\right)  \vy \\
&= \vv^{T}\left(r\right) \left(\mV^{T} \mV \right)^{-1} \mV^{T}  \mX^{T}  \va  \left(\theta\right)  \\
&= \vv^{T}\left(r\right) \mU \mX^{T} \va \left(\theta\right) ,
\label{equ_temp}
\end{split}
\end{align}
where $\mU = \left(\mV^{T} \mV \right)^{-1} \mV^{T}$. Computing \eqref{equ_temp} involves matrix vector products for each tuple. Having an analytical formula also permits us to obtain the overall area average
\begin{equation}
T_{avge} =  \frac{1}{\pi\left(r_{o}^{2}-r_{i}^{2}\right)}\int_{r_{i}}^{r_{o}}  \int_{0}^{2 \pi}   T\left(r,\theta\right)r drd\theta,
\label{equ_expectation}
\end{equation}
%and the variance 
%\begin{equation}
%\mathbb{\sigma}^{2}\left[T\left(r,\theta\right)\right]=\int\int T^{2}\left(r,\theta\right)r\rho\left(r,\theta\right)drd\theta- \left( E\left[T\left(r,\theta\right)\right] \right)^2,
%\label{equ_variance}
%\end{equation}
%both implicitly conditioned upon the measurement values in $\mB$. 
where $r_o$ and $r_i$ are the outer and inner radii respectively. This formula will be used later when computing area averages given measurements $\mB$. Computing the standard root-mean-squared error $\varepsilon_{p}$ (with respect to the measurements) is also straightforward and given by
\begin{align}
\begin{split}
\varepsilon_{p}^{2} &=\frac{1}{NM}\left\Vert \mA \mX- \mB\right\Vert _{2}^{2} \\
&=\frac{1}{NM} \text{vec}\left(\mB \right)^{T}  \left( \mI_{M} \otimes \left(  \mI_{N} - \mQ \mQ^{T} \right) \right) \text{vec}\left(\mB \right),
\label{equ:errors}
\end{split}
\end{align}
where one can utilize the thin QR factorization of $\mA$, i.e.~$\mA=\mQ \mR$, and $\mR^{T} \mX = \mQ^{T} \mB$ for ease in computation. In \eqref{equ:errors} the symbol $\otimes$ denotes the Kronecker product and $\text{vec}$ denotes a vectorized version of a matrix, obtained by stacking columns of the matrix sequentially. 

\subsection{Frequency selection algorithm}
This error metric can be used to frame the following optimization problem
\begin{align}
\begin{split}
\underset{  \boldsymbol{\omega} \in \mathcal{I}, \; \mX \in \mathbb{R}^{\left(2k+1 \right) \times M } }{\text{minimize}} \; \; & \varepsilon_{p}\left( \mX, \boldsymbol{\omega}  \right) \\
\text{subject to} \; \; &  \left\Vert \mX \right\Vert _{2}^{2}  \leq \beta^{2}.
\end{split}
\label{equ:opt}
\end{align}
This requires us to iterate over the space of integers $\mathcal{I}$ for obtaining the frequencies, while simultaneously iterating over the space of $\left(2k+1 \right) \times M$ real matrices---making \eqref{equ:opt} a challenging non-convex optimization problem to solve. Further, solutions have to satisfy the inequality constraint where $\beta$ is a constant tailored to avoid solutions with large norms. We expand upon this salient point below.

For some combinations of $\boldsymbol{\omega}$ and rake positions $\boldsymbol{\theta}$, $\mA$ can be ill-conditioned, implying that the linear equations are \emph{undetermined} \cite{hansen2010discrete}. Standard least squares solves on such matrices result in $\mX$ having a large norm, manifesting in our problem as large amplitude overshoots between measurement points. This would result in Fourier series expansions that closely match the temperature values at prescribed $\boldsymbol{\theta}$, but give wildly varying temperature values at other circumferential locations. As such overshoots are not physical and entirely numerical, we \emph{regularize} the problem with the addition of the inequality constraint in \eqref{equ:opt}. 

\begin{figure}[h]
\centering
\includegraphics[scale=0.5]{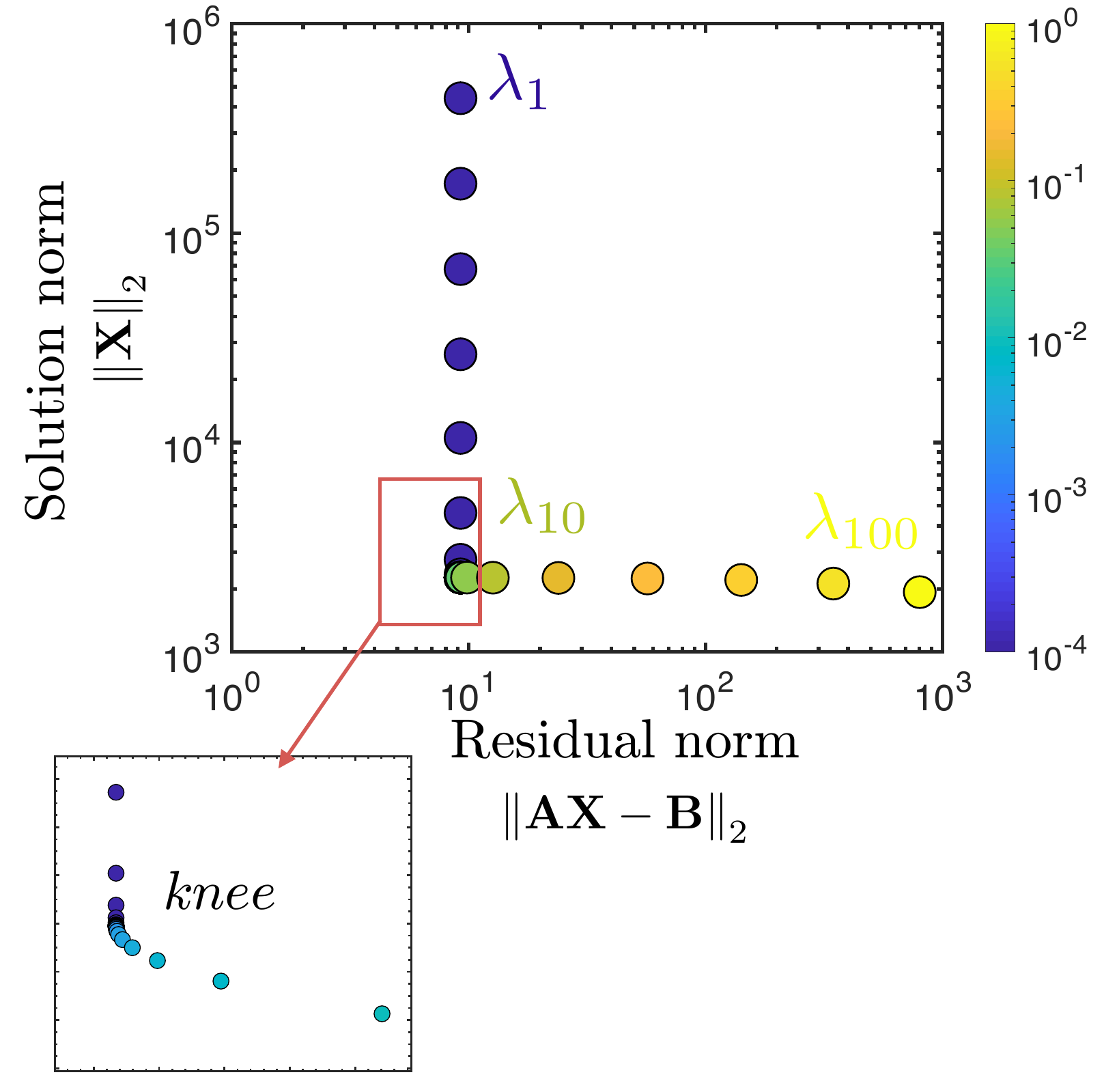}
\caption{The L-curve for the regularized least squares problem on an extract from Engine A with $\boldsymbol{\omega} = \left(2, 5 \right)$.}
\label{fig:lcurve}
\end{figure}

While iterative optimization strategies can be constructed to solve \eqref{equ:opt}, we pursue a different approach. We first select a suitable set of frequencies $\boldsymbol{\omega}$, construct $\mA$ and then find $\mX$ by solving the \emph{regularized} least squares problem
\begin{align}
\underset{\mX}{\text{minimize}} \; \; & \left\Vert \mA\mX-\mB\right\Vert _{2}^{2} + \left\Vert \lambda  \mX \right\Vert _{2}^{2}.
\label{equ:opt_constrained_A}
\end{align}
Here $\lambda$ is a scalar value, chosen such that one obtains a favorable compromise between a sufficiently smooth solution and at the same time a small residual. A well-worn criterion for selecting $\lambda$ is based on finding the \emph{knee} of the \emph{L-curve} (see 4.7 in \cite{hansen2010discrete})---a log-log scatter plot of the solution norm on the horizontal axis and the residual norm on the vertical axis, plotted for different values of $\lambda$. A representative example is shown in Fig.~\ref{fig:lcurve} for an extract from Engine A with $\boldsymbol{\omega} = \left(2, 5 \right)$. Each marker here denotes the solution norm and the residual norm for a particular choice of $\lambda$. The range of $\lambda$ values used in our studies (and in the plot in Fig.~\ref{fig:lcurve}) was varied from $10^{-10}$ to $1$. Non-graphical methods for finding $\lambda$ can be found in \cite{castellanos2002triangle}.

Once an appropriate value for $\lambda$ has been chosen, one solves for $\mX$ using
 \begin{equation}
\mX =  \left( \mA^{T} \mA + \lambda^2 \mI \right)^{-1} \mA^{T} \mB.
 \end{equation}
We encode this approach into our overall brute force frequency selection strategy, shown in Algorithm \ref{alg:bruteforce}. The idea here is to iterate over the different frequency combinations that yield low values of the RMS error. Our choice of the number of harmonics is dictated largely by the number of rakes in Engines A, B, C and D (6 in all of them), which permits $\mA$ to have a maximum of 6 columns. We therefore restrict ourselves to two frequencies, $k=2$, which results in $\mA$ having dimensions of $6 \times 5$. The maximum value of the frequency pairs is also restricted to focus on lower harmonics, i.e.~$\text{max}\left\{ \boldsymbol{\omega}  \right\} \leq 10$.

%The resulting \emph{reguarlized} least squares problem takes the form
%\begin{align}
%\begin{split}
%\underset{\mX}{minimize} \; \; & \left\Vert \mA\mX-\mB\right\Vert _{2}^{2} \\
%\text{subject to} \; \; &   \left\Vert \mX \right\Vert _{2}^{2}  \leq \beta
%\end{split}
%\end{align}
%Applying Lagrange multipliers to 
%\begin{equation}
%\underset{X}{minimize}\left\Vert \left(\begin{array}{c}
%\mA\\
%\lambda \mI
%\end{array}\right) \mX-\left(\begin{array}{c}
%\mB\\
%\beta \mE
%\end{array}\right)\right\Vert _{2}^{2}
%\end{equation}
\begin{algorithm}
\caption{Brute force frequency selection for two harmonics.}
\label{alg:bruteforce}
\begin{algorithmic}[1]
\STATE{Set $\boldsymbol{\omega}=\left(\omega_1, \omega_2 \right)$, where $\omega_1 \neq \omega_2$ and $\text{max}\left\{ \boldsymbol{\omega}  \right\} \leq 10$.}
\STATE{Solve $\hat{\mX}={\text{argmin}}\left\Vert \mA\mX-\mB\right\Vert _{2}^{2}$}
\STATE{Set $\boldsymbol{\lambda} = \left(0.0001, 0.001, 0.1, 10 \right)$}
\WHILE{$\left\Vert \hat{\mX} \right\Vert _{2}\geq\beta$}
\STATE{$\lambda_i = \boldsymbol{\lambda}\left( i \right)$}
\STATE{Solve $\hat{\mX}={\text{argmin}}\left\Vert \mA\mX-\mB\right\Vert _{2}^{2} + \left\Vert \lambda_{i} \mX \right\Vert _{2}^{2}$  }
\ENDWHILE
\STATE{Compute $\varepsilon_{p}^{2} = \left\Vert \mA \hat{\mX} -\mB\right\Vert _{2}^{2} / NM$}
\RETURN $\varepsilon_{p}^{2}$ \end{algorithmic} \end{algorithm}

\begin{figure}
\centering
\includegraphics[scale=0.5]{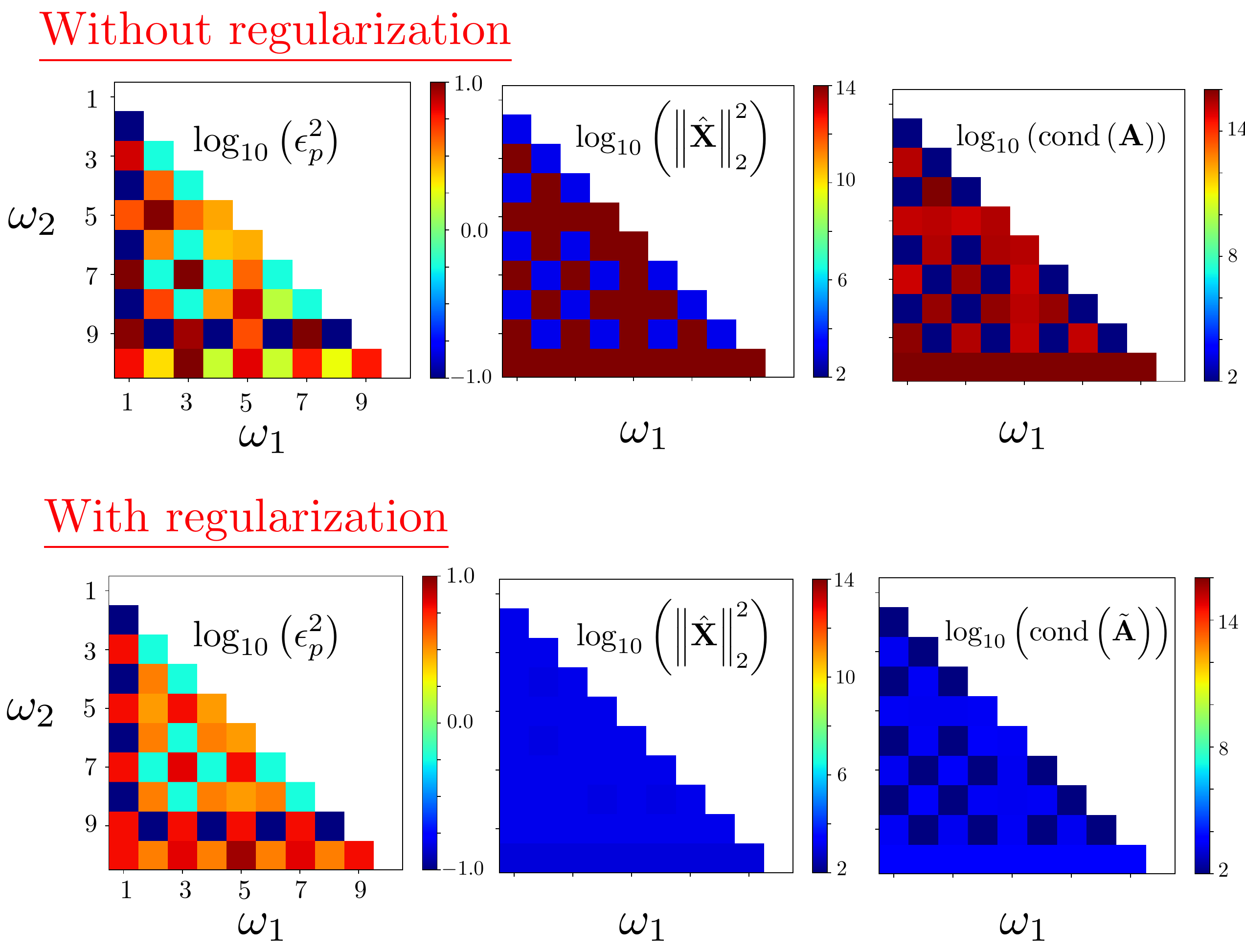}
\caption{Brute force frequency selection with and without regularization for the last extract from the Engine A data-set.}
\label{fig:regularization}
\end{figure}

In Algorithm \ref{alg:bruteforce} we have made the assumption that regularization is necessary. To further examine this assumption, consider the results shown in Fig.~\ref{fig:regularization} both with and without (commenting out lines 4 to 7 in Algorithm \ref{alg:bruteforce}) regularization. Shown here are the values for $\varepsilon_{p}^{2}$, $\left\Vert  \hat{\mX} \right\Vert _{2}^{2}$ and the condition number of $\mA$ (all on a base-10 logarithmic scale) for various frequency pairs. For the regularized case, we plot the condition number of the \emph{augmented system} given by 
\begin{equation}
\tilde{\mA} = \left(\begin{array}{c}
\mA\\
\lambda^2 \mI
\end{array}\right).
\end{equation} 

It should be clear from Fig.~\ref{fig:regularization} that, although regularization does increase $\varepsilon_{p}$ for certain frequency pairs, it offers a more stable system of linear equations, resulting in physically plausible temperature variations. We reiterate the importance of having solutions with small norms, as they are unlikely to exhibit massive oscillations between measurement points, i.e., they will be more smooth. In other words, it is better to have solutions that have non-zero residuals and are smooth, than solutions that have a residual of zero and are non-smooth. 

\subsection{Identifying suitable harmonics}
\begin{figure}
\centering
\includegraphics[scale=0.5]{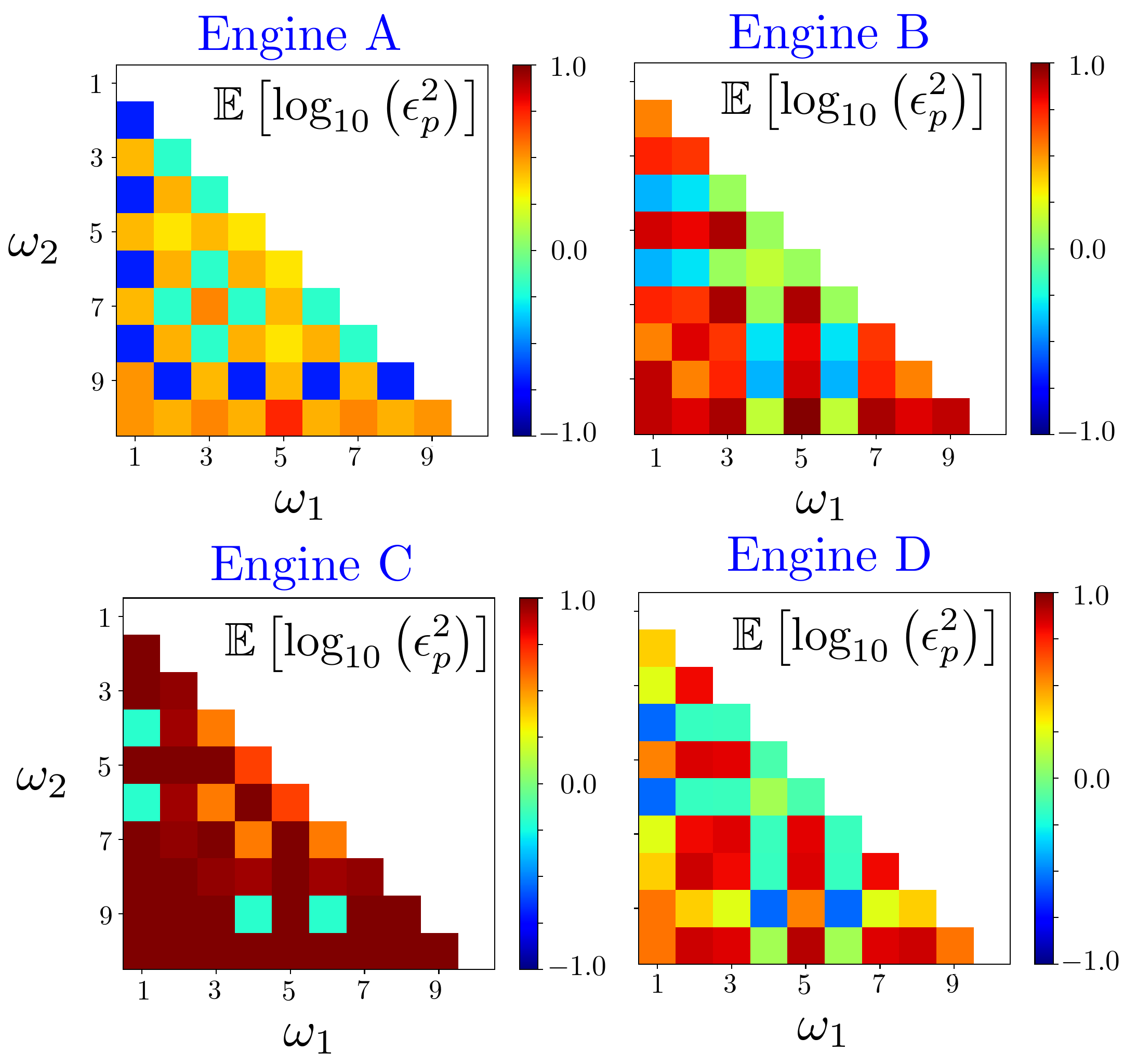}
\caption{Average values of the RMS error when applying Algorithm 1 to the data-sets of Engines A, B, C and D.}
\label{fig:rms_errors}
\end{figure}

We apply Algorithm \ref{alg:bruteforce} to the 70 extracts in Engines A, B, C and D with $\beta=10^5$. Fig. \ref{fig:rms_errors} plots the average values of $\varepsilon_{p}$ for the different frequency pairs. It is apparent from these results that across all these extracts---corresponding to a range of different operating points on each engine's power curve---there are four frequency pairs that consistently yield low errors: $\boldsymbol{\omega}=(1,4)$, $\boldsymbol{\omega}=(1,6)$, $\boldsymbol{\omega}=(4,9)$ and $\boldsymbol{\omega}=(6,9)$. 

\begin{figure}
\centering
\includegraphics[scale=0.45]{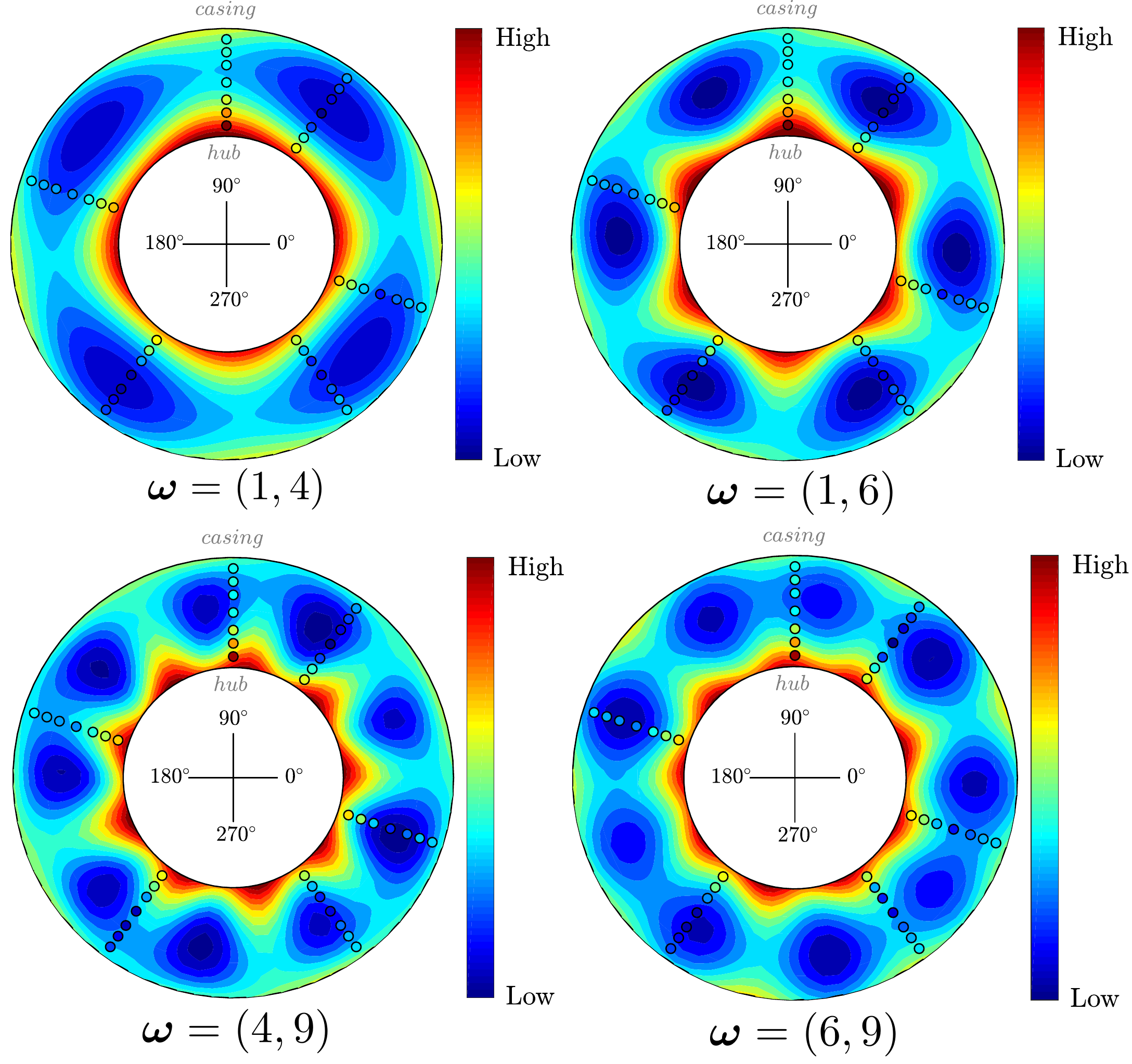}
\caption{Spatial representations of the temperature shown for the first extract in Engine A. The contour bounds are set by the minimum and maximum temperature measurements.}
\label{fig:spatial_fields}
\end{figure}

Full spatial representations of temperature associated with these harmonics are shown in Fig.~\ref{fig:spatial_fields} for the first extract of Engine A. The contours are generated using the Fourier series expansion in the circumferential direction, based on the values of $\hat{\mX}$, and a quadratic polynomial-based extrapolation in the radial direction, as explained in Sec.~\ref{sec:model}. The colored circular markers shown in the four subfigures of Fig.~\ref{fig:spatial_fields} denote the thermocouple positions on the six rakes and their corresponding temperature values. In what follows, we focus our attention solely on the four harmonic pairs identified above. 

Before we move on, it will be worthwhile to discuss \emph{aliasing} in the context of our frequency selection algorithm. Aliasing (see page 91 in \cite{strang1996wavelets}) is an artifact that causes signals to be indistinguishable when sampled; typically the original signal is sampled at an insufficient number of points to re-create it in its entirety. In our context, based on the placement of the rakes and the frequencies selected for approximation, one can get multiple frequencies that interpolate the sampled data points exactly. 

\subsection{Training and testing the model}
Thus far we have not distinguished between \emph{training} and \emph{testing} data-sets---a key practice in data-centric approaches. One typically infers the parameters of a model on the training data, and tests its suitability on the testing data. Bootstrap or $K-$fold cross-validation are widely adopted (see Chapter 7 in \cite{friedman2009elements}). With regard to the latter, caution must be exercised when selecting the number of folds: too few and the cross-validation estimator can have high variance as the training data is similar; too many and we risk incurring a bias. In what follows, we train and test our multivariate regression model on Engine E. Recall our prior experiments on Engines A, B, C and D have pointed us towards four particular harmonic pairs. Here we aim to further prune down the number of harmonics that best estimate the spatial temperature field across all the engines\footnote{Assuming that the same engine modes are prevalent over different engines measured at the same measurement plane (relative to upstream and downstream components).}.

\begin{figure}[t]
\centering
\includegraphics[scale=0.5]{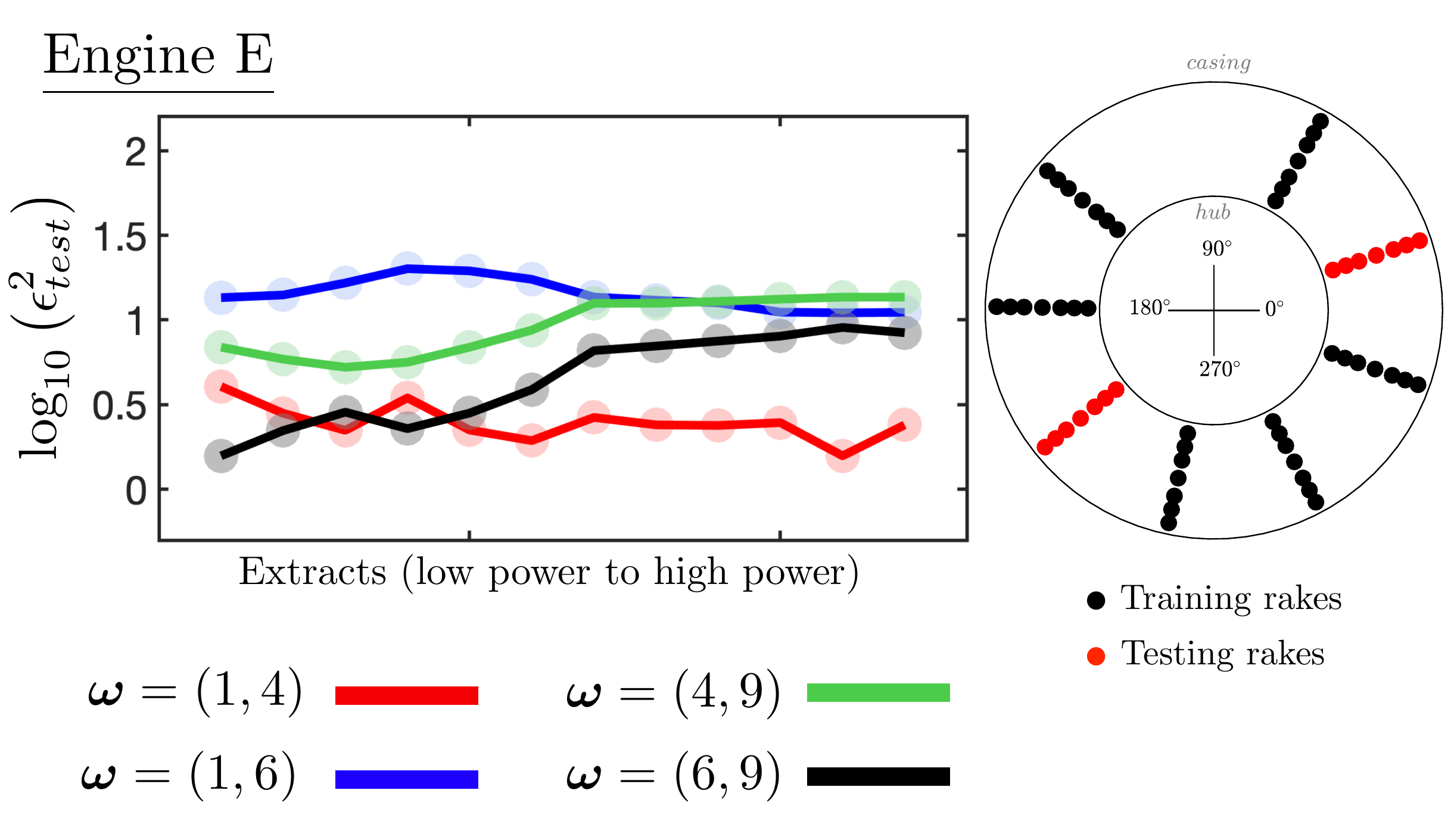}
\caption{Testing the multivariate regression model on Engine E with certain rakes reserved for training and the remainder for testing.}
\label{fig:testing_frequencies}
\end{figure}

We utilize a \emph{leave P out} cross-validation on Engine E; in the case of Engine E, we leave out two rakes for testing while six rakes are used for training. While we note that this engine has more rakes than Engines A, B, C and D, we wish to avoid over-fitting the data, and thus restrict ourselves to using six rakes for training. For clarity, we denote the position of the circumferential rakes used for training by $\boldsymbol{\theta}_{train}$ and those for testing by $\boldsymbol{\theta}_{test}$. Definitions of the measurements follow suit, with $\mB_{test} \in \mathbb{R}^{N_{test} \times M}$ and $\mB_{train} \in \mathbb{R}^{N_{train} \times M}$. Finally, the training Fourier matrix is specified by 
\begin{equation}
\mA_{train} = \mA \left( \boldsymbol{\omega}, \boldsymbol{\theta}_{train} \right) \in \mathbb{R}^{N_{train} \times \left(2k+1 \right) }
\end{equation}
and the testing one by 
\begin{equation}
\mA_{test} = \mA \left( \boldsymbol{\omega}, \boldsymbol{\theta}_{test} \right) \in \mathbb{R}^{N_{test} \times \left(2k+1 \right) }
\end{equation}
where we set $N_{train}=6$ and $N_{test}$ to be the remaining rakes. In cross-validation our objective is to minimize the predictive error, given by
\begin{equation}
\varepsilon_{test}^{2}  = \frac{1}{NM} \left\Vert {\mA}_{test} \hat{\mX} -\mB_{test} \right\Vert _{2}^{2}.
\end{equation}
Consider the results in Fig.~\ref{fig:testing_frequencies} shown for a single combination of training and testing rakes, chosen randomly and repeated for the four different frequency pairs. The plots show the predictive error as a function of the various extracts. 

\begin{figure}[h]
\centering
\includegraphics[scale=0.7]{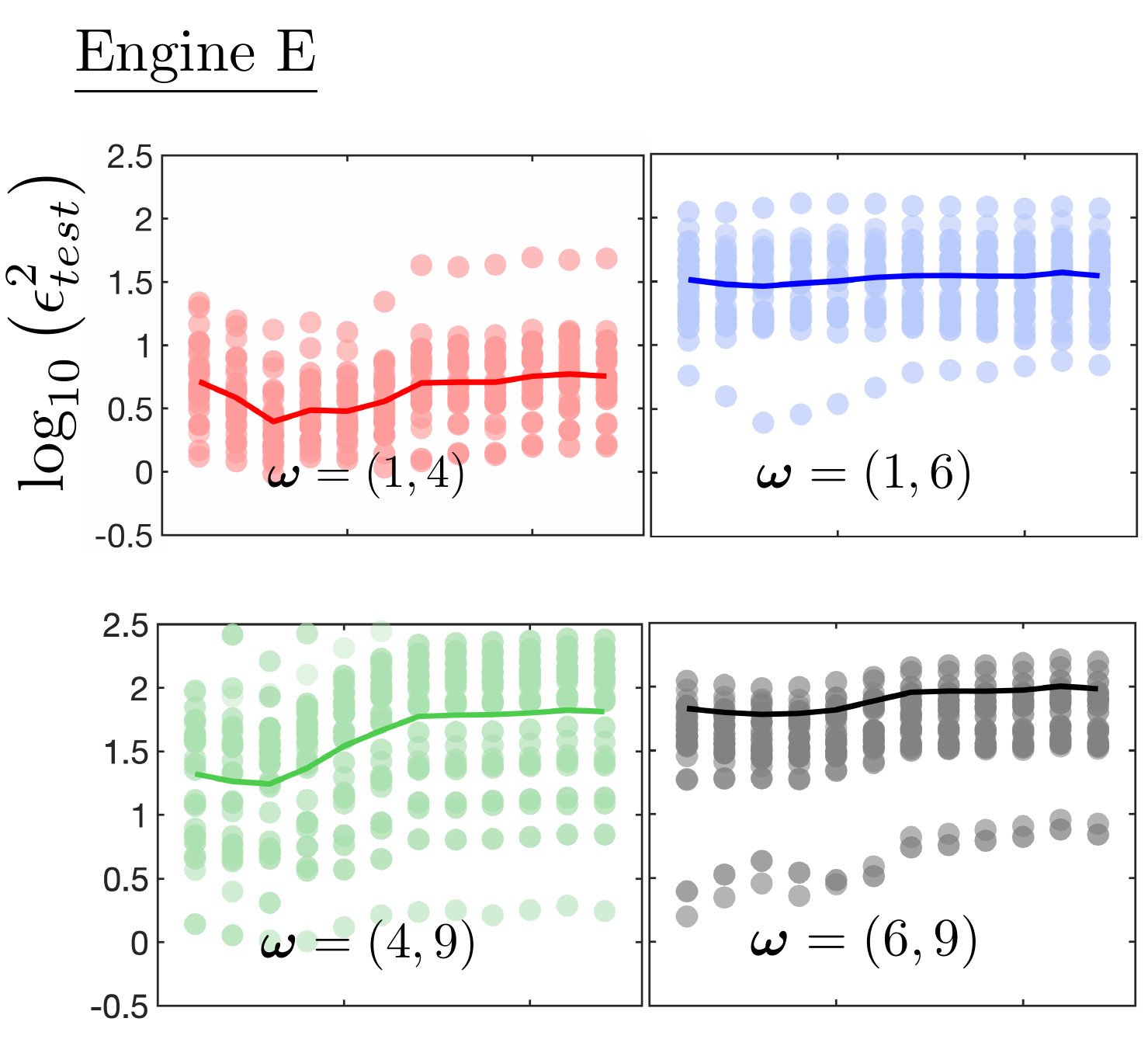}
\caption{Cross-validation results with 28 rake combinations; predictive errors over the various extracts are shown as a function of the four frequency pairs.}
\label{fig:crossvalidation}
\end{figure}

Given that we have eight potential rake locations and we restrict ourselves to selecting only six for training (and the remaining for testing), we have 28 possible combinations. For each of these 28 rake arrangements, we compute the errors; the results are shown in Fig.~\ref{fig:crossvalidation}. The shaded transparent markers show the predictive errors for the different trials (rake arrangements), while the thick lines denote the mean values of $\varepsilon_{test}^{2}$ over the 28 trials. Note, it is clear from these results that on average $\boldsymbol{\omega}=\left(1, 4\right)$ yields the lowest value of $\varepsilon_{test}^{2}$, both in expectation and in variance---the latter shown by the reduced scatter in the results. 

\section{An assumed case study}
\label{sec:area}
In section~\ref{sec:lit} we mentioned that owing to the constraints imposed by the Nyquist-Shannon sampling theorem, we cannot recover the amplitudes and phases of all the harmonics using classical fast Fourier transform techniques. We also commented that, although methods based on $l_1$ norm minimization can potentially aid in signal recovery, they require sparsity in the coefficients. In this section, we study two particular aspects of these statements with the objective of recovering the entire spatial temperature profile---not simply the first two harmonics---with sparse measurements. 

\begin{figure}
\centering
\includegraphics[scale=0.6]{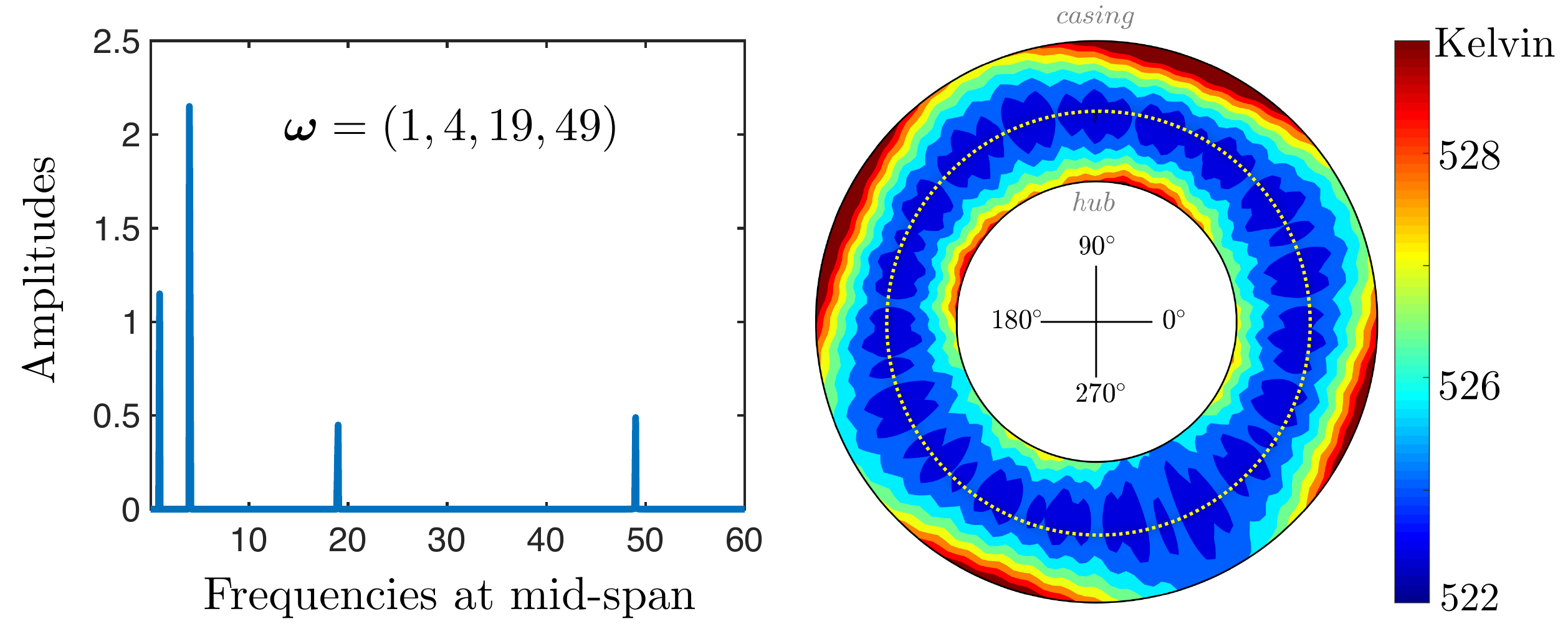}
\caption{An assumed temperature profile with four harmonics; the amplitude and phases are varied from hub to casing.}
\label{fig:assumed_profile}
\end{figure}

In our analysis thus far, we have operated under the premise that we do not know what the \emph{true} spatial variation in temperature is. To expose our methods, and to offer techniques to the wider turbomachinery community, we study an analytically generated temperature profile, shown in Fig.~\ref{fig:assumed_profile}. This profile has an average temperature value of 526.85 K and is comprised of four harmonics $\boldsymbol{\omega}=\left(1, 4, 19, 49 \right)$. 

\subsection{Identifying the harmonics}
Our goal here is apply the regularized least squares approach from Sec.~\ref{sec:model} on this temperature profile to determine whether the approach can capture the two dominant frequencies with only 6 rakes; in some sense serving as a validation of Algorithm 1. 

\begin{figure}
\centering
\includegraphics[scale=0.6]{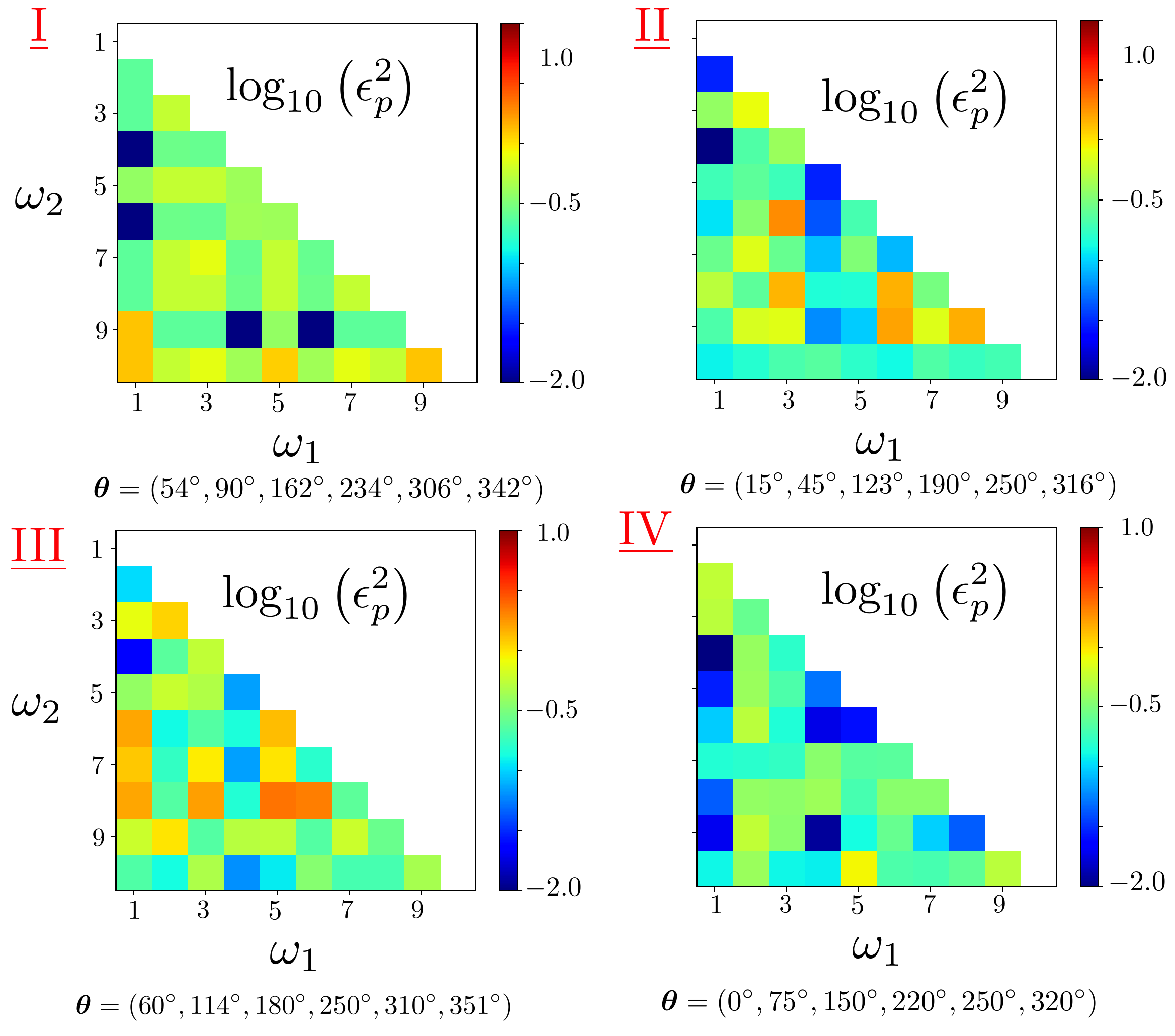}
\caption{Values of the error $\varepsilon^{2}$ for four different sets of rake positions.}
\label{fig:error_assumed}
\end{figure}

We study four different sets of virtual rake positions, defined by the values of $\boldsymbol{\theta}$, shown in Table~\ref{table:rakes}. The first is based on the arrangement in Engine A, while the other three are randomly selected such that there is at least one rake in each of the four quadrants. Fig.~\ref{fig:error_assumed} plots the $\varepsilon_{p}^{2}$ (once again on a base-10 logarithm scale) errors using all 6 rakes---i.e., we do not split the data into a testing and training set. It is clear that across all four rake positions, $\boldsymbol{\omega} = \left(1, 4 \right)$ does yield the lowest error; however, there are other frequency pairs that also yield low errors, e.g., $\boldsymbol{\omega} = \left(1, 6 \right)$, $\boldsymbol{\omega} = \left(4, 9 \right)$ and $\boldsymbol{\omega} = \left(6, 9 \right)$, but only for some sets of rake positions. 

\begin{table}
\begin{center}
%\resizebox{\columnwidth}{0.95}{
\caption{Rake positions for the assumed temperature study.}
\begin{tabular}{|c|c|}
  \hline		
 Case & Rake positions (in deg$^\circ$) \\ \hline
 \hline
I & 54, 90, 162, 234, 306, 342 \\
II & 15, 45, 123, 190, 250, 316 \\
III & 60, 114, 180, 250, 310, 351 \\
IV & 0, 75, 150, 220, 250, 320 \\
  \hline  
\end{tabular}
\label{table:rakes}
\end{center}
\end{table}

\begin{figure}
\centering
\includegraphics[scale=0.53]{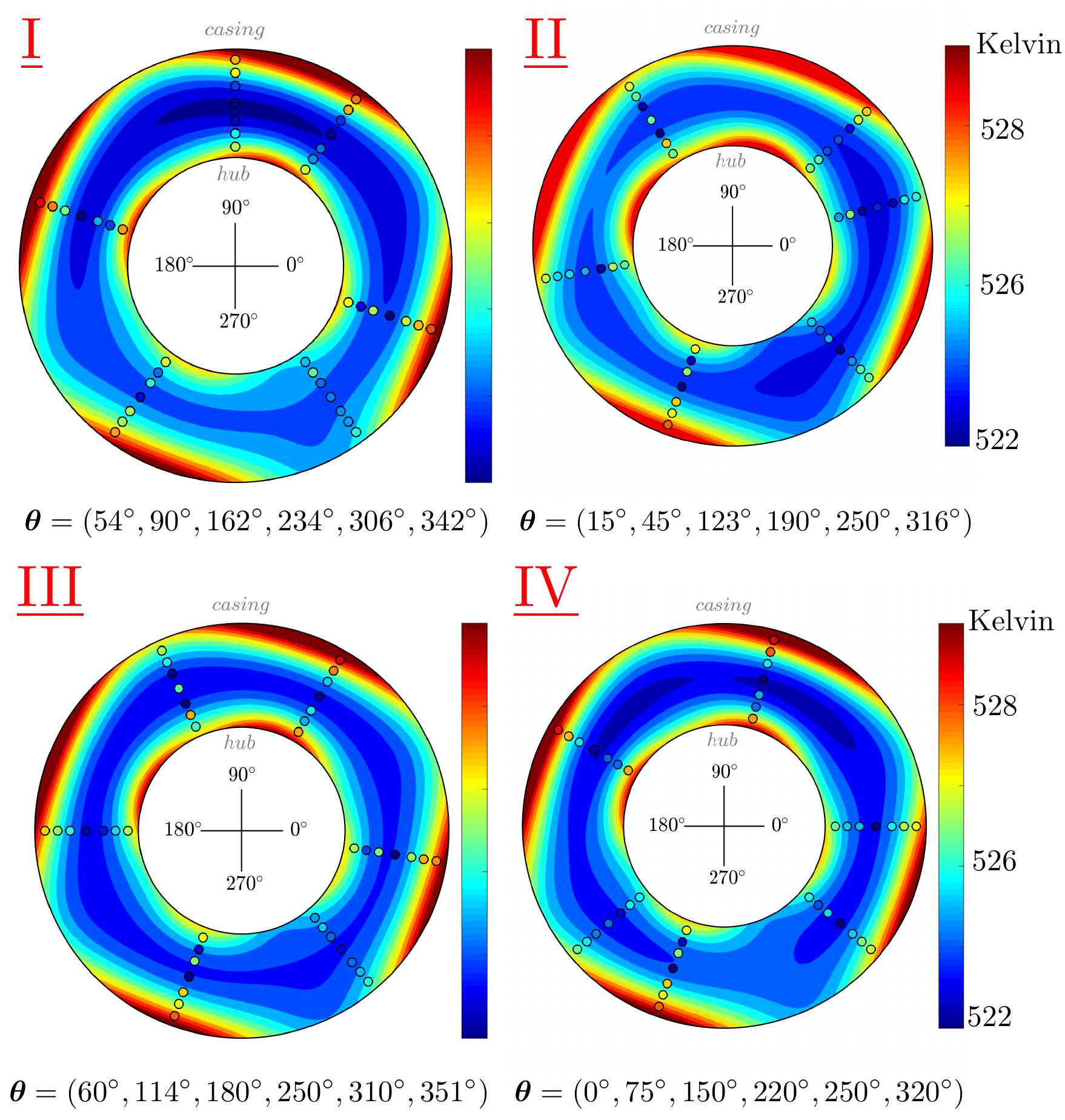}
\caption{Temperature profiles for four different rake positions with $\boldsymbol{\omega}=\left(1, 4 \right)$.}
\label{fig:final_temps}
\end{figure}

The temperature profiles associated with $\boldsymbol{\omega} = \left(1, 4 \right)$ are illustrated in Fig.~\ref{fig:final_temps} for the four different rake positions. Although the rake positions are different, the spatial profiles are similar. In fact, even if the rakes were circumferentially equidistant from each other, one can still recover similar patterns to Fig.~\ref{fig:final_temps}.

\subsection{Signal approximation}
\begin{figure}
\centering
\includegraphics[width=8.0cm]{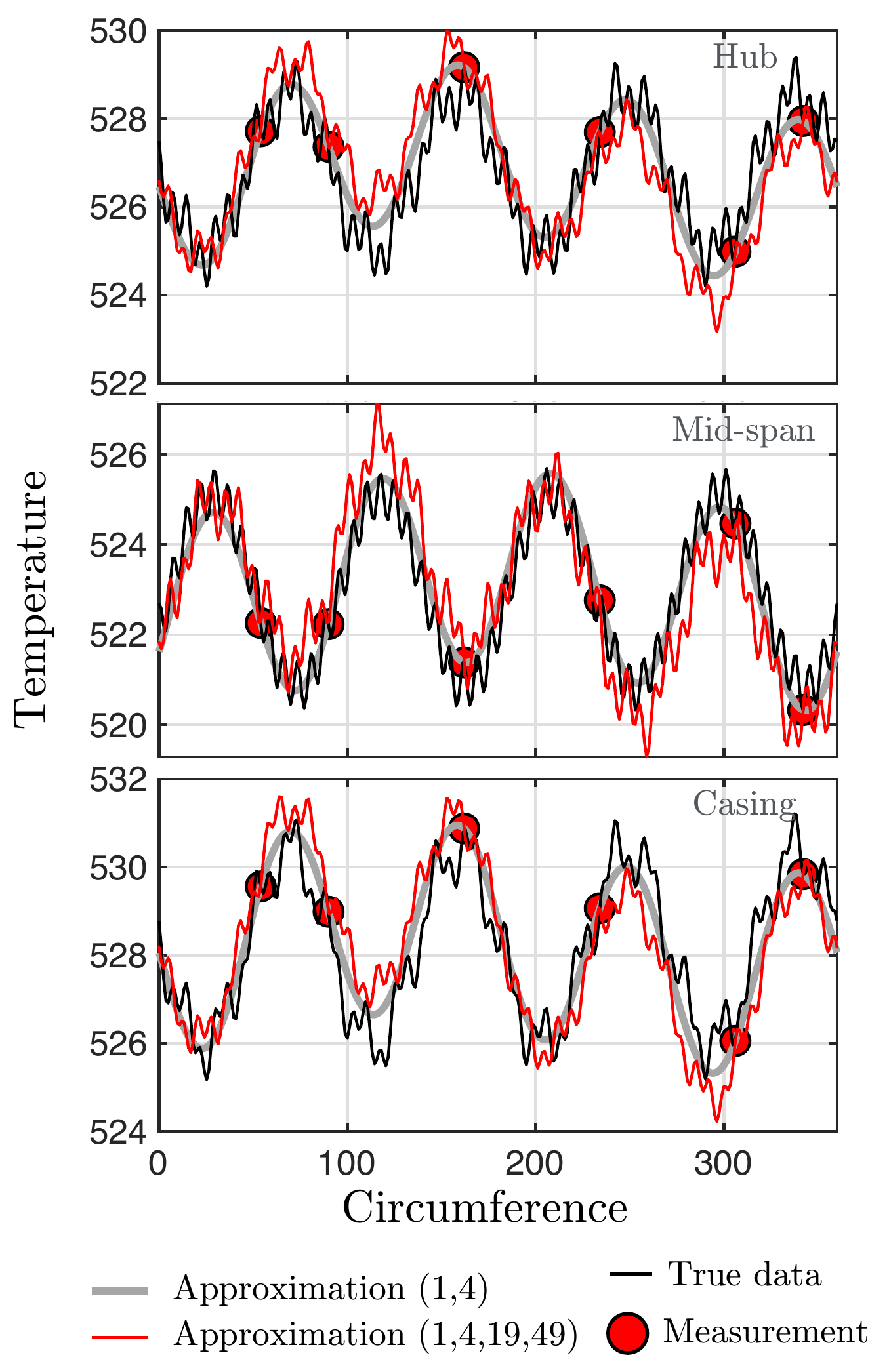}
\caption{An assumed temperature profile with four harmonics; the amplitude and phases are varied from hub to casing.}
\label{fig:assumed_profile_slices}
\end{figure}

In this subsection, we are interested in ascertaining whether we can estimate the amplitudes and phases of all four harmonics given 6 rakes, assuming a priori information on the harmonics is known. This leads to the solution of an undetermined system of equations where $\mA \in \mathbb{R}^{6 \times 9}$. While there are infinitely many solutions to such a linear system, we are interested in the solution that has the lowest $l_2$ norm. Note that this solution strategy is different from the previously discussed L-curve approach, which is typically used on tall or square matrices; our matrix here is fat with more columns than rows. One approach to compute the \emph{minimum norm} solution is given below. Let 
\begin{equation}
\mA \left[\begin{array}{cc}
\mP_{1} & \mP_{2}\end{array}\right] = \left[\begin{array}{cc}
\mQ_{11} & \mQ_{12}\end{array}\right] \left[\begin{array}{cc}
\mR_{11} & \mR_{12}\\
\boldsymbol{0} & 0
\end{array}\right]
\end{equation}
be the pivoted QR factorization\footnote{The pivoted QR factorization is a well-known heuristic for subset selection. See page 276 in Golub and Van Loan \cite{golub2013matrix}} of $\mA$, where $\mP$ is a permutation matrix, $\mQ \in \mathbb{R}^{6 \times 6}$ is an orthogonal matrix and $\mR \in \mathbb{R}^{6 \times 9}$ has the sub-matrix $\mathbf{R}_{11}$ that is upper triangular. These matrices have been partitioned based on the numerical rank of $\mA$, which depends on the rake positions in $\boldsymbol{\theta}$. In this example we use rake positions corresponding to case I, which yields a rank of 5. Thus, $\mR_{11} \in \mathbb{R}^{5 \times 5}$ (and is invertible), $\mP_{1} \in \mathbb{R}^{9 \times 5}$ and $\mQ_{11} \in \mathbb{R}^{6 \times 5}$. The \emph{least norm} solution is then given by
\begin{equation}
\mX = \mP_{1} \mR_{11}^{-1} \mQ_{11}^{T} \mB,
\label{equ:least_norm}
\end{equation}
where as before $\mB \in \mathbb{R}^{6 \times 7}$. Fig. \ref{fig:assumed_profile_slices} plots the circumferential distribution of temperature at measurement points close to the hub, mid-span and casing. The red line in this figure corresponds to the least norm solution (obtained using \eqref{equ:least_norm}), the thick gray line corresponds to the standard least squares approximation with only the first two harmonics, and the black line represents the true temperature distribution.

While the least norm solution does not offer perfect signal recovery, it is able to match the measurements exactly at the measurement locations, and offers an acceptable approximation to the true temperature distribution. It should be noted that this heuristic works even if we alter the amplitudes of the four harmonics---even setting them to be equivalent.

\subsection{A comparison of area averages}
It is worth reiterating our motivation for approximating the full spatial temperature field. First, images such as those shown in Fig.~\ref{fig:final_temps} are useful to understand key engine modes and to identify spatial asymmetries. They are also very important for computing averages.

\begin{figure}
\centering
\includegraphics[width=8.5cm]{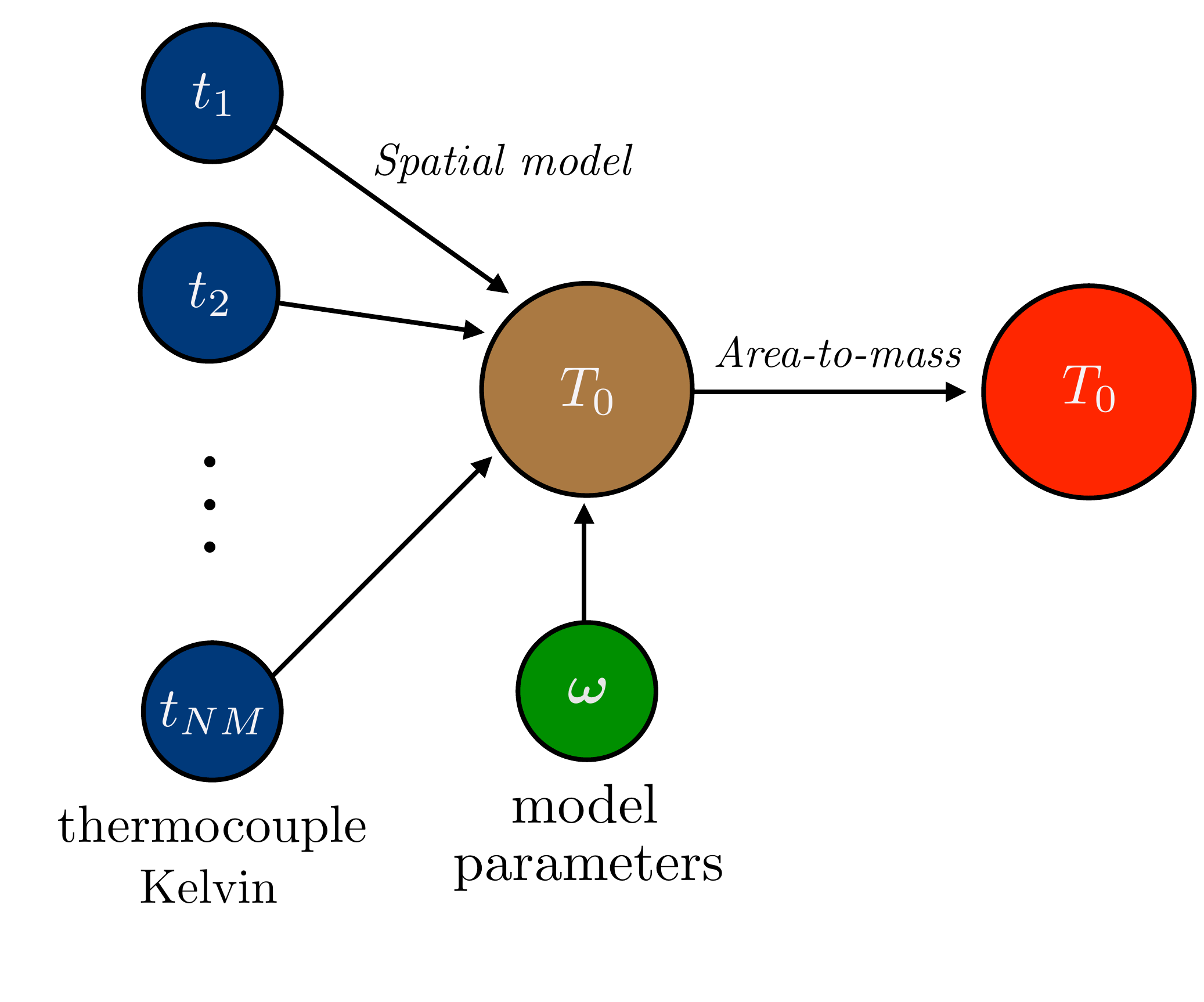}
\caption{Workflow from thermocouple measurements to a spatial average.}
\label{fig:workflow}
\end{figure}

Recall the discussion in section~\ref{sec:intro} on averaging practices. For the engine measurements considered throughout this paper, we do not have mass-flow rate distributions to compute mass-weighted or momentum-weighted averages. It is not uncommon to area average the temperature measurements and then compute an area-to-mass conversion factor, as shown in Fig. \ref{fig:workflow}. This conversion is usually estimated by CFD on the engine component; a process that in itself has uncertainties well beyond the scope of this work. What we are concerned with here is the way the area average in Fig.~\ref{fig:workflow} is computed. 

When using the standard area averaging approach, each measurement is weighted by the sector area coverage associated with its probe (see page 8 in \cite{skiles1980turbine}). Thus, no allowance is made to account for scenarios where the rake positions may capture only the peaks or the troughs of the wave forms. In fact, for the assumed profile considered in this section, the area weighted average technique yields a temperature of 526.20 K. 

For integrating the approximated spatial field, consider the formula in \eqref{equ_expectation}. This can be written as
\begin{align}
\begin{split}
T_{avge} =  & = \frac{1}{\pi\left(r_{o}^{2}-r_{i}^{2}\right)}\int_{r_{i}}^{r_{o}}  \int_{0}^{2 \pi} \vv^{T} \left(r\right) \mU \mX^{T} \va \left(\theta\right)   r dr d\theta \\
& =  \frac{2}{\left(r_{o}^{2}-r_{i}^{2}\right)}\int_{r_{i}}^{r_{o}}  r \vv^{T} \left(r\right) \mU\left[\begin{array}{cc}
\vx_{1}^{T} & \boldsymbol{0}
\end{array}\right] dr
\end{split}
\label{equ:area_new}
\end{align}
where we have used the fact that the harmonic terms in $\va$ (all terms except the first one), when integrated between 0 to $2\pi$, will become zero. Here $\vx_{1}^{T}$ corresponds to the first column of $\mX^{T}$, which comprises of the constant terms of each of the $M$ Fourier expansions. As both approximations in Fig. \ref{fig:assumed_profile_slices}, corresponding to $\boldsymbol{\omega}=(1,4)$ and $\boldsymbol{\omega} = (1,4,19,49)$, are able to obtain the same value of the constant terms, their area averages computed using \eqref{equ:area_new} are equivalent. We report an area average of 525.85 K, which is greater than the area weighted value, but equivalent to the profile average temperature value. Area average temperature comparisons across extracts in Engines A through E---between \eqref{equ:area_new} and the sector weighted area average value---revealed differences between 0.5 and 2 K.

\section*{Conclusions}
In this paper, we have developed a data-centric model for engine temperature measurements. Our model takes as an input the temperature values obtained from a few circumferentially placed rakes and outputs a 2D spatial temperature field. A key component of our strategy is an iterative approach for selecting frequency pairs (restricted in our investigation to the first two harmonics due to limits on the number of rakes) that introduces regularization to avoid solutions with large norms---manifesting as large overshoots between successive data points. 

There is a compelling case to be made regarding area averaging using our strategy. Rather than computing an average solely based on temperature values and their positions, our framework fits a 2D spatial model to the data to estimate its average. The model need not necessarily capture all the spatial harmonics, but in some cases it can still deliver the true area average. Our investigation in this paper has also revealed the importance of rake positions and the assumed temperature values themselves. Future work will be aimed at studying the impact of uncertainties in these measurements.

\section*{Acknowledgements}
The authors are grateful to Ra\'{u}l V\'{a}zquez D\'{i}az (Rolls-Royce) and John Longley (Cambridge University). This work was funded by Rolls-Royce plc; the authors are grateful to Rolls-Royce for permission to publish this paper. 

\bibliography{references}

\begin{thebibliography}{10}

\bibitem{ernst2011analysis}
Ernst, M., Michel, A., and Jeschke, P., 2011.
\newblock ``Analysis of rotor-stator-interaction and blade-to-blade
  measurements in a two stage axial flow compressor''.
\newblock {\em Journal of Turbomachinery, \textbf{ 133}}(1), p.~011027.

\bibitem{sanders2002multi}
Sanders, A., Papalia, J., and Fleeter, S., 2002.
\newblock ``Multi-blade row interactions in a transonic axial compressor {Part
  I: Stator} particle image velocimetry ({PIV}) investigation''.
\newblock {\em Journal of Turbomachinery, \textbf{ 124}}, pp.~10--18.

\bibitem{mailach2008periodical}
Mailach, R., Lehmann, I., and Vogeler, K., 2008.
\newblock ``Periodical unsteady flow within a rotor blade row of an axial
  compressor {Part I: Flow} field at midspan''.
\newblock {\em Journal of Turbomachinery, \textbf{ 130}}(4), p.~041004.

\bibitem{stoll1979effect}
Stoll, F., Tremback, J.~W., and Arnaiz, H.~H., 1979.
\newblock {Effect of Number of Probes and their Orientation on the Calculation
  of Several Compressor Face Distortion Descriptors}.
\newblock Tech. rep., NASA Technical Memorandum 72859.

\bibitem{francis1989measurement}
Francis, S.~T., and Morse, I.~E., 1989.
\newblock {\em {Measurement and Instrumentation in Engineering: Principles and
  Basic Laboratory Experiments}}, Vol.~67.
\newblock CRC Press, Boca Raton, FL.

\bibitem{cumpsty2006averaging}
Cumpsty, N.~A., and Horlock, J.~H., 2006.
\newblock ``Averaging nonuniform flow for a purpose''.
\newblock {\em Journal of Turbomachinery, \textbf{ 128}}(1), pp.~120--129.

\bibitem{pianko1983propulsion}
Pianko, M., and Wazelt, F., 1983.
\newblock {Suitable Averaging Techniques in Non-Uniform Internal Flows}.
\newblock Tech. rep., AGARD-AR-182.

\bibitem{dzung1971}
Dzung, L.~S., 1971.
\newblock ``Consistent mean values for compressible media in the theory of
  turbomachines''.
\newblock {\em Brown Boveri Review, \textbf{ 58}}(10), pp.~485--492.

\bibitem{seshadri2019b}
Seshadri, P., , Duncan, A., Simpson, D., Thorne, G., and Parks, G., 2019.
\newblock ``Spatial flow-field approximation using few thermodynamic
  measurements {Part II: Uncertainty} assessments''.
\newblock {\em Submitted to ASME Journal of Turbomachinery}.

\bibitem{skiles1980turbine}
Skiles, T.~W., 1980.
\newblock {Turbine Engine Flowpath Averaging Techniques}.
\newblock Tech. rep., Arnold Air Force Station, Texas.

\bibitem{chilla2019reducing}
Chilla, M., Pullan, G., and Gallimore, S., 2019.
\newblock ``Reducing instrumentation errors caused by circumferential flow
  field variation in multi-stage axial compressors''.
\newblock In ASME Turbo Expo 2019: Turbomachinery Technical Conference and
  Exposition, American Society of Mechanical Engineers, pp.~1--11.

\bibitem{friedman2009elements}
Hastie, T., Tibshirani, R., and Friedman, J., 2009.
\newblock {\em The Elements of Statistical Learning}, 2nd~ed.
\newblock Series in Statistics. Springer, New York, NY.

\bibitem{rogers2016first}
Rogers, S., and Girolami, M., 2016.
\newblock {\em A First Course in Machine Learning}.
\newblock CRC Press, Boca Raton, FL.

\bibitem{goodfellow2016deep}
Goodfellow, I., Bengio, Y., and Courville, A., 2016.
\newblock {\em Deep Learning}.
\newblock MIT Press, Cambridge, MA.

\bibitem{hansen2010discrete}
Hansen, P.~C., 2010.
\newblock {\em Discrete Inverse Problems: Insight and Algorithms}.
\newblock Fundamentals of Algorithms. SIAM, Philadelphia, PA.

\bibitem{castellanos2002triangle}
Castellanos, J.~L., G{\'o}mez, S., and Guerra, V., 2002.
\newblock ``The triangle method for finding the corner of the {L-curve}''.
\newblock {\em Applied Numerical Mathematics, \textbf{ 43}}(4), pp.~359--373.

\bibitem{strang1996wavelets}
Strang, G., and Nguyen, T., 1996.
\newblock {\em {Wavelets and Filter banks}}.
\newblock SIAM, Philadelphia, PA.

\bibitem{golub2013matrix}
Golub, G.~H., and Van~Loan, C.~F., 2013.
\newblock {\em {Matrix Computations}}, Vol.~4.
\newblock JHU Press, Baltimore, MD.

\end{thebibliography}
\bibliographystyle{asmems4}
\end{document}